\documentstyle[aps,twocolumn]{revtex}
\newcommand{\Label}[1]{\label{#1}}

\begin{document}

\title{A particle-number-conserving Bogoliubov method which demonstrates the 
validity of the time-dependent Gross-Pitaevskii equation for a highly condensed 
Bose gas\\
\ \ \\
{\small \sl Final version 27 February 1997}}
\author{C. W. Gardiner}
\address{Physics Department, Victoria University of Wellington, Wellington, New 
Zealand}

\maketitle
\begin{abstract}
The Bogoliubov method for the excitation spectrum of a Bose-condensed gas is 
generalized to apply to a gas with an exact large number $ N$ of particles.  
This 
generalization yields a description of the Schr\"odinger picture field 
operators as the {\em product} of an annihilation operator $ A$ for the total 
number of particles and the sum of a ``condensate wavefunction'' 
$ \xi({\bf x})$ 
and a {\em phonon}  field operator $ \chi({\bf x})$ in the form 
$ \psi({\bf x}) \approx A\{\xi({\bf x}) + \chi({\bf x})/\sqrt{N}\}$  when the 
field operator acts on the $ N $ particle subspace.  It is then 
possible to expand the Hamiltonian in decreasing powers of $\sqrt{N}$, 
an thus obtain solutions for eigenvalues and eigenstates
as an asymptotic expansion of the same kind.  It is also possible to 
compute all matrix elements of field operators between states of {\em 
different} $N$.

The excitation spectrum can be obtained by essentially the same method as 
Bogoliubov only if $ \xi({\bf x})$ is a solution of the time independent 
Gross-Pitaevskii equation for $ N$ particles and any chemical potential $ \mu$
which yields a valid and stable solution of the Gross-Pitaevskii equation.  
The treatment within a subspace of fixed $N$ is identical in form to 
that usually used, but the interpretation of the operators is slightly 
different.

A time-dependent generalization is then made, yielding an asymptotic 
expansion
in decreasing powers of $ \sqrt N$ for the equations of motion.
In this expansion the condensate wavefunction has the time-dependent form 
$ \xi({\bf x},t)$, and the condition for the validity of the expansion is that 
$ \xi({\bf x},t) $ satisfies the time-dependent Gross-Pitaevskii equation
${\partial\xi /\partial t }= -({\hbar^2/ 2m})\nabla^2\xi 
+V\xi + N u  |\xi |^2 \xi$.

The physics is then described in a kind of interaction picture, called the {\em 
condensate picture}, in which the phonon operator can be expressed as
$ \chi({\bf x},t) = \sum_k\xi_k({\bf x},t)\alpha_k$, where the operators
$ \alpha_k$ are {\em time independent} annihilation operators, and the 
state-vector has a time evolution described by a Schr\"odinger equation
in which the Hamiltonian is a time-dependent quadratic form in the phonon 
creation and annihilation operators, whose coefficients are explicitly 
determined in terms of the time-dependent condensate wavefunction 
$ \xi({\bf x},t)$
\end{abstract}

\pacs{PACS Nos.  03.75.Fi%BEC coherent
,05.30.Jp%boson
,51.10.+y%kinetics of gases
,05.30.-d%Qstatmech
}

%\begin{multicols}{2}
\section{Introduction}\Label{sec1}
Two central tools in the description of a Bose-condensed gas are the Bogoliubov 
method \cite{Bogoliubov,Abrikosov,LandP,Fetter} and the 
Gross-Pitaevskii equation \cite{GPoldies} in both its time independent 
and its time dependent guises.  Because Bose condensates 
\cite{CondensateExperiments} can now be be experimentally produced, 
there has been renewed interest in both of these tools
\cite{Lewenstein,Fetter1996,Lewenstein et al,Lewenstein You,Javanainen 1/97,%
Burnett Nist}
\cite{Burnett95a,Burnett95b,Ballagh,Edwards1996a,Edwards1996b,Baym,%
Stringari96a,Stringari96b,Stringari96c,YouHolland,HollandCooper}.

As it is presently formulated, Bogoliubov's method treats the 
condensate operators a c-numbers. One consequence of this is that
the resulting approximate Hamiltonian does not conserve the total 
number of particles. The enforcement of number conservation in the 
mean only leads to a description which is essentially confined to a 
subspace with a single value of the mean number of particles.

Another difficulty is that, although there are derivations of the 
validity of the time independent Gross-Pitaevskii equation as a 
description of the condensate ground state wavefunction (these 
appear largely as consequences of the adaptation of the Bogoliubov 
method to a trapped condensate), there does not appear to be any 
derivation of the validity of the time-dependent Gross-Pitaevskii 
equation as a description of the motion of a trapped condensate.

This paper will solve both of these problems and show that their 
solutions are strongly connected. 

We will firstly show how to modify Bogoliubov's argument in such a way 
as not to break the conservation of particle number.  This 
modification yields a description of the particle field operator $ 
\psi({\bf x})$ in the form
\begin{eqnarray}\Label{I1}
 \psi({\bf x}) \approx
 A\left(\xi({\bf x}) + {1\over\sqrt{N}} \chi({\bf x})\right)
\end{eqnarray}
where $ A$ is an annihilation operator such that the eigenvalue of $ 
A^\dagger A$ is $ N$, the total number of particles, $ \xi({\bf x})$ 
is the condensate wavefunction, and $ \chi({\bf x})$ is a phonon field 
operator.  The Bogoliubov Hamiltonian is expressed in terms of these 
phonon operators, and the non-conservation of phonons which arises is 
not unexpected.  The eigenstates of this Bogoliubov Hamiltonian are 
nevertheless all states with exactly $ N $ particles, so there is no 
doubt that {\em particle} number conservation is not violated.  The 
method is approximate, of course, but the accuracy of the 
approximations made is exactly the same as that of the usual 
Bogoliubov method.

It is clear that the separation of the {\em phonon} concept from the 
{\em particle} concept that enables the method to succeed.  To emphasize the 
validity of the the phonon concept we show that our expression for 
$ \psi({\bf x})$ can be used to demonstrate that the quantized phase of 
$ \psi({\bf x})$ is the velocity potential operator, and that the operator of 
density fluctuations is its canonical conjugate, as is well known 
\cite{LandP}.

The treatment of the spatially inhomogeneous case requires more care 
in defining exactly what the expansion procedure is.  The Bogoliubov 
method and the Gross-Pitaevskii equations are both used to describe 
the weakly interacting Bose gas; the requirement that the gas is 
weakly interacting is formalized in our treatment by requiring the 
interaction potential $ u$ to be of order of magnitude $ 1/N$; that 
is, we write $ u = \tilde u/N$, and carry out the asymptotic expansion 
in decreasing powers of $ \sqrt{N}$ at fixed $ \tilde u$.

The treatment is then similar for the case of the time independent or 
the time dependent condensates.  The formula (\ref{I1}) is substituted 
in the Hamiltonian, and the terms of different degree in $ \sqrt N$ 
grouped together yielding
\begin{eqnarray}\Label{I2}
H = {N}\,{\cal H}_1 + \sqrt{N}\,{\cal H}_2 + {\cal H}_3
\end{eqnarray}
$ {\cal H}_1 $ is found to be a c-number, $ {\cal H}_2$ is linear in phonon 
operators and $ {\cal H}_3$ is quadratic in phonon operators.  

In the time-independent case the ground state is found by minimizing $ 
{\cal H}_1$, and this means that $ \xi({\bf x},t)$ must satisfy the 
time-independent Gross-Pitaevskii equation---this also means that the 
linear term $ {\cal H}_2$ vanishes, and the excitation spectrum is 
given by diagonalizing the quadratic part $ {\cal H}_3$.

The description of the inhomogeneous condensate that results is 
almost isomorphic to that originally used by Fetter \cite{Fetter1972} 
when restricted to a subspace of definite $N$, and thus is in 
agreement with other recent calculations.  However, we obtain in 
addition a description of the relationship between the states for 
different $N$, and in particular are able to present the matrix 
elements of the field operators between states of $N$ and $N+1$ 
particles.  (It was the need to have these in a description of 
condensate growth which motivated this work originally.)

In the time-dependent case the logic is slightly different.  In 
general there is an explicit time-dependence proportional to $ \sqrt N 
$ in $ \chi({\bf x},t)$ as a result of the representation of the 
time-independent $ \psi({\bf x})$ operator in terms of a time 
dependent condensate wavefunction $ \xi({\bf x},t)$, and this being of 
order of magnitude $ \sqrt N $ contradicts the asymptotic 
representation in the form (\ref{I1}).  The part proportional to $ 
\sqrt N $ however can be made to cancel with the implicit time 
dependence arising for $ \sqrt N {\cal H}_2$ provided $ \xi({\bf 
x},t)$ satisfies the time dependent Gross-Pitaevskii equation: {\em 
thus it is only possible to get an asymptotic expansion of the form 
(\ref{I1}) if $ \xi({\bf x},t)$ is a solution of the time dependent 
Gross-Pitaevskii equation.} The remaining time dependence can adjusted 
by a suitable choice phonon mode functions $ \xi_k(t)$ to yield a 
representation in terms of time independent phonon annihilation 
operators, and by going into an appropriate interaction picture---the 
{\em condensate picture}---the time dependence of the state vector is 
given by a Schr\"odinger equation $ {\cal H}_3(t)|\Phi,t\rangle = 
i\hbar d |\Phi,t\rangle/ dt$.  Since the number of atoms is fixed, 
this gives a full description of the system; the condensate 
wavefunction satisfies the time-dependent Gross-Pitaevskii equation, 
and the motion of the residual quantized phonons is given by $ {\cal 
H}_3(t)$

\section{A number conserving Bogoliubov transformation for the homogeneous Bose 
gas}\Label{sec2}
Let us start with the problem of the weakly interacting Bose gas confined in a 
large box, with no explicit trapping potential.  The Hamiltonian in momentum 
space is written
\begin{eqnarray}\Label{n1}
H &=& \sum_{\bf k}\hbar\omega_{\bf k}a^{\dagger}_{\bf k}a_{\bf k}
\nonumber \\ &&
+{u\over 2}\sum_{{\bf k}_1,{\bf k}_2,{\bf k}_3,{\bf k}_4}
a^{\dagger}_{{\bf k}_1}a^{\dagger}_{{\bf k}_2}a_{{\bf k}_3}a_{{\bf k}_4}
\delta_{{\bf k}_1+{\bf k}_2,{\bf k}_3+{\bf k}_4}.
\end{eqnarray}
The basis states are usually written
\begin{eqnarray}\Label{1}
|n_0,{\bf n}\rangle
\end{eqnarray}
where $ {\bf n}\equiv \{n_{\bf k}\} $ represents the vector of occupation 
numbers of all particles with non-zero momentum, and the dependence on $ n_0$ 
is explicitly separated.
Now let us write them in a form which eliminates reference
to $ n_0$ by using the total number $ N = n_0 +\sum_{\bf k} n_{\bf k}$,
in the form
\begin{eqnarray}\Label{2}
|N,{\bf n}\rangle
\end{eqnarray}
We define the operators $ A$, $ {\cal N}$ and $ \alpha_{\bf k}$ as follows
\begin{eqnarray}\Label{3}
A|N,{\bf n}\rangle&=& \sqrt{N}|N-1,{\bf n}\rangle\\
{\cal N}  &\equiv& A^\dagger A \\
 \alpha_{\bf k} &=& {1\over\sqrt{\cal N}}a_0^\dagger a_{\bf k}  	
\end{eqnarray}
The operators $ \alpha_{\bf k}$ are essentially phonon operators, while $ A$
and $ {\cal N}$ are operators which refer to the total numbers of particles.  
Since the phonon operators $ \alpha_{\bf k}$ commute with the total number 
operator $ {\cal N}$, we can make approximations involving the phonon 
operators without violating the conservation of total numbers of particles.
\subsubsection{Relationship between $ a_0$ and  $ A$}
The operator $ A $ reduces the total number of particles by 1, without changing 
the number of $ {\bf k}\ne 0$ particles; it is thus proportional to the 
operator $ a_0$.  In fact the precise relationship is
\begin{eqnarray}\Label{3001}
a_0 |N,{\bf n}\rangle&=&\sqrt{n_0}|N-1,{\bf n}\rangle
\\&=& \sqrt{1-{{\sum_{{\bf k}\ne 0} n_{\bf k}\over N}}}\,\,
 A |N,{\bf n}\rangle
\end{eqnarray}
If $ N $ is very large compared with 
$ \sum_{{\bf k}\ne 0} n_{\bf k}$---that is, the system is Bose 
condensed---it is clear that we may make the approximation
\begin{eqnarray}\Label{4}
a_0  |N,{\bf n}\rangle
&=& A\left({1-{{\sum_{{\bf k}\ne 0} n_{\bf k}\over {2 N}}}}\right)
  |N,{\bf n}\rangle
\\ \Label{4a}
&\approx & A |N,{\bf n}\rangle .
\end{eqnarray}
In what follows we will be expanding in inverse powers of $ N^{1/2}$, 
and it will sometimes  be necessary to keep the more accurate form (\ref{4}).

\subsubsection{Relationship between $ a_{\bf k}$ and  $ \alpha_{\bf k}$}
One can similarly show that one can write
\begin{eqnarray}\Label{401}
a_{\bf k}|N,{\bf n}\rangle&\approx & 
{A\alpha_{\bf k}\over \sqrt{N+1-\sum_{\bf k}n_{\bf k}}}|N,{\bf n}\rangle
\\
 &\approx& {1\over\sqrt{ N}}
\left( 1 +
 {\sum_{{\bf k}\ne 0}n_{\bf k}-1\over 2 N}
\right)
  A\alpha_{\bf k}|N,{\bf n}\rangle
\\ \Label{401a}
&\approx& A{\alpha_{\bf k}\over\sqrt{ N}}|N,{\bf n}\rangle;
\end{eqnarray}
however we will not find it necessary to use anything other than the simple 
form (\ref{401a}).
\subsubsection{Commutators of the $ \alpha_{\bf k}$}
When acting on such condensed states with large $ N$, we also find that we can 
approximate the commutator 
\begin{eqnarray}\Label{5}
[\alpha_{\bf k},\alpha_{{\bf k}'}^\dagger] &=&
 \delta_{{\bf k}{\bf k}'} -{1\over N}a_{\bf k}a^\dagger_{{\bf k}'}\\
&\approx& \delta_{{\bf k}{\bf k}'}.
\end{eqnarray}
\subsubsection{Validity of the approximations}
The approximations made are  accurate to the same order in $ N$ as those 
usually assumed in the Bogoliubov theory.  For clarity we have written 
equations such as (\ref{3},\ref{3001},\ref{401}) etc., as explicitly acting on 
a state with $ N $ particles.  However in the remainder of the paper we will 
simply write equalities between operators, which are understood to be valid 
when 
these operators act on a vector in the subspace of $ N$ particles.

\subsection{Transformation of the Hamiltonian}
The Hamiltonian (\ref{n1}) is now approximated firstly, as is usual in 
Bogoliubov theory by dropping the 
terms in the interaction which do not involve at least two operators 
$ a_0$, $a_0^\dagger$, and then secondly by making 
the replacements (and appropriate Hermitian conjugates)
\begin{eqnarray}\Label{6}
a^\dagger_0a_{\bf k} &\to & \alpha_{\bf k}\sqrt{\cal N}  \\
a^\dagger_0a_0  &\to & {\cal N} \\
a^\dagger_{\bf k} a_{\bf k} &\to & \alpha^\dagger_{\bf k} \alpha_{\bf k}
\end{eqnarray}
in the Hamiltonian.  Notice that for all of these replacements the left hand 
sides have essentially the same action on a number state as the right hand 
sides, but with a 
modified coefficient; that is for all of them the non-zero matrix elements 
occur for the same states, and the coefficients are almost equal, in contrast 
to the usual Bogoliubov transformation, which changes the essential nature of 
the operators in the number state basis.

As in the usual Bogoliubov method, the term involving 
$ a_0^\dagger a_0^\dagger a_0a_0 $ has to be treated more accurately, 
effectively by using the more accurate approximation (\ref{4}), giving
\begin{eqnarray}\Label{301}
a_0^\dagger a_0^\dagger a_0a_0 
&=&  a_0^\dagger a_0a_0^\dagger a_0  - a_0^\dagger a_0
\nonumber \\
&\approx & {\cal N}^2 - {\cal N} 
        - 2{\cal N}\sum_{\bf k}a^\dagger_{\bf k}a_{\bf k}
\nonumber \\
&\approx &{\cal N}^2 - {\cal N} 
        - 2{\cal N}\sum_{\bf k}\alpha^\dagger_{\bf k}\alpha_{\bf k}.
\end{eqnarray}
Although in this case it is usual to set $ \omega_{ 0} = 0$, for consistency 
with the inhomogeneous case we do not make this assumption, and this requires a 
similar correction to the term $ \hbar\omega_{ 0}a^\dagger_0 a_0$.
The  Hamiltonian (\ref{n1}) then becomes 
\begin{eqnarray}\Label{7}
H & =& {1\over 2}u\left({\cal N}^2 -{\cal N}\right)
+\sum_{{\bf k}\ne 0}\hbar(\omega_{\bf k} - \omega_{0})
 \alpha^\dagger_{\bf k}\alpha_{\bf k}
\nonumber \\
&& + 
{u{\cal N}\over2}\sum_{{\bf k}\ne 0}
\left\{2\alpha^\dagger_{\bf k}\alpha_{\bf k}+ \alpha_{\bf k}\alpha_{-{\bf k}}
+ \alpha^\dagger_{\bf k}\alpha^\dagger_{-{\bf k}}\right\}
\end{eqnarray}
It is to be understood that this operator form is valid on all states, 
including superpositions of states with different numbers of particles, as long 
as only states with large eigenvalues of $ {\cal N}$ are included.
The Hamiltonian can be diagonalized for any eigenvalue of $ {\cal N}$, to get 
the usual Bogoliubov spectrum; the eigenstates are simultaneous eigenstates 
of $ {\cal N}$ and $ H$.  
\subsubsection{Asymptotic expansion}
The expression of the Hamiltonian in the form (\ref{7}) can be put in the form 
of an asymptotic expansion in inverse powers of $ {N}$ in the sense that the 
restriction of $ H$ to a subspace of definite $ N$ is approximated by (\ref{7}) 
for sufficiently large $ N$, provided it is understood that $ u$ is itself a 
small quantity of order of magnitude $ 1/N$.  Under that condition it can be 
checked that  approximations made in its derivation affect only the terms of 
lower order in $ N $; i.e., the corrections are of order $ 1/N$. 
The requirement that $ u $ be ``small'' amounts to an assumption that the 
kinetic plus trapping potential terms, and interaction energy terms in 
(\ref{7}), are of the same order of 
magnitude.  A formal statement of this requirement can be made by 
writing $ u = \tilde u/N$ in (\ref{7}), and then solving the problem by making 
an expansion in appropriate decreasing powers of $ N $.

The Hamiltonian in this form has two major advantages over the deceptively 
similar conventional form.  The most significant is that the conservation of 
total numbers of particles is maintained; the quasiparticles on the other hand 
appear only as phonons, and are not conserved.  The second major advantage is 
that the approximation method is a systematic expansion in inverse powers of 
the large quantity $ N $, which has a definite value in any case one considers.

\subsubsection{Expression of field operators in terms of the velocity potential 
operator}
The consequences of this method on the representation of the field operators 
are interesting.  We can show that the representation of phase and density 
fluctuations as in \cite{LandP} is a natural consequence.

The field operator is 
\begin{eqnarray}\Label{8}
\psi({ {\bf x}}) &=&
{1\over\sqrt{V}}\left( a_0 + \sum_{{\bf k}\ne 0} a_{\bf k}  e^{i{{\bf k}\cdot {
\bf 
x}}}
\right)
\\ & \approx & 
{A\over\sqrt{V}}\left(1 + {1\over\sqrt{ N}}\sum_{{\bf k}\ne 0}\alpha_{\bf k}  
e^{i{{\bf k}\cdot {\bf x}}}\right)
\end{eqnarray}
We now introduce the Bogoliubov transformation, approximated by assuming 
$ {\bf k}$ is 
very small (though this is not an essential assumption, and is introduced only 
to make it match up with \cite{LandP}), as
\begin{eqnarray}\Label{9}
\alpha_{\bf k} = {b_{\bf k} - b^\dagger_{-{\bf k}}\over\sqrt{2\hbar k/m v(N)}}
   	   	+\sqrt{\hbar k\over 2mv(N)} b^\dagger_{-{\bf k}}
\end{eqnarray}
where $ v(N) = \sqrt{uN/m}$ is the speed of sound for long wavelengths.
Clearly, the second term is much smaller than the first for small $ {\bf k}$, 
so we keep only the first part initially.  We then find that
\begin{eqnarray}\Label{10}
\psi({ {\bf x}}) &\approx& 
{A\over\sqrt{V}}\left[1 + \sum_{{\bf k}\ne 0} \sqrt{mv\over 2\hbar k{ N}}(
b_{\bf k}  e^{i{{\bf k}\cdot {\bf x}}}-b^\dagger_{\bf k}  e^{-i{{\bf k}\cdot {
\bf x}}}
)\right]\!\!\!\!\!
\\
&\approx& {A\over\sqrt{V}} \exp\big(i\Phi({\bf x})\big)
\end{eqnarray}
since
\begin{eqnarray}\Label{11}
\Phi({\bf x}) = -i \sum_{{\bf k}\ne 0} \sqrt{mv\over 2\hbar k{ N}}\left\{
b_{\bf k}  e^{i{{\bf k}\cdot {\bf x}}}-b^\dagger_{\bf k}  e^{-i{{\bf k}\cdot {
\bf x}}}
\right
\}
\end{eqnarray}
is small.

This matches up exactly with (27.1) of \cite{LandP}.  That is, the operator $ 
\Phi$ is 
the quantized velocity potential operator, with the substitution
 $ b_{\bf k} \to ic_{\bf k}$ 
to match up with their notation. If we now include the next order term we find 
density fluctuations as well.  
Using the full form (\ref{9}) for the density
\begin{eqnarray}\Label{12}
\psi^\dagger({\bf x})\psi({\bf x}) &\approx&
{{ N}\over V}
 + {1\over V}\sum_{{\bf k}\ne 0} \sqrt{{ N}\hbar k\over 2mv }
\left\{b_{\bf k}  e^{i{{\bf k}\cdot {\bf x}}} + b^\dagger_{\bf k} 
 e^{-i{{\bf k}\cdot {\bf x}}}\right\}
\nonumber\\ &&
\\
&=& \rho_0 + {\delta\rho({\bf x})}
\end{eqnarray}
giving the correct operator for the density fluctuations, as in \cite{LandP} 
(24.10).

In this limit that only long wavelengths are involved, the Hamiltonian 
(\ref{7}) can be rewritten in terms of a part related to sound 
waves, as in \cite{LandP}, and an additional purely $ {\cal N}$ dependent part:
\begin{eqnarray}\Label{sound}
H & =& {1\over 2}u\left({\cal N}^2 -{\cal N}\right) +E_0({\cal N})
\nonumber\\
&&+{1\over 2}:\int d^3{\bf x}\,\bigg\{
\rho_0\big(\nabla\cdot\Phi({\bf x})\big)^2
+v({\cal N})^2{\delta\rho({\bf x})^2\over\rho_0}\bigg\}: \, .
\nonumber\\
\end{eqnarray}
Here $ E_0({\cal N}) $ is the Bogoliubov ground state energy.  

Notice that
not only do $ \rho_0$ and $ v({ N}) $ depend on $ { N}$, but that the 
very definitions of the $ \Phi({\bf x})$ and $ \delta\rho({\bf x}) $ also 
depend on $ { N}$.  However, if we believe that a superselection rule 
applies whereby only eigenstates of $ { N}$ occur in nature, the correct 
treatment of states with uncertain total numbers of particles must arise from 
an incoherent superposition of solutions, for various $ N $, of the equations 
of motion arising from (\ref{sound}).

\section{The spatially inhomogeneous case}\Label{sec3}
The spatially inhomogeneous situation, which can arise either by the existence 
of 
vortices or because there is a trapping potential confining the gas, was first 
treated by Fetter 
\cite{Fetter1972}, and in the case of a trapped gas, has been treated more 
recently in \cite{Fetter1996,Lewenstein You,Javanainen 1/97,%
Burnett Nist}.

The standard formalism, as used by Fetter \cite{Fetter1972}, is based on the 
work of Hugenholtz and Pines \cite{Hugen} which in summary, shows that the 
correct ground state and excitation spectrum is given by the following 
prescription:
\begin{enumerate}
\item Replace $ a_0^\dagger \to \sqrt{N}$, $ a_0 \to \sqrt{N}$ in both
$ H $ and $ {\cal N}$.
\item For a given $ \mu$, find the value of $ N$ and the state which minimize
$\langle K\rangle $, where $ K = H - \mu N$.  This gives the ground state 
energy.
\item The excited states are given by the higher eigenstates of $ K $ with 
$ a_0^\dagger \to \sqrt{N}$, $ a_0 \to \sqrt{N}$.  By neglecting terms in $ N$
of order less than $ N$, one obtains the Bogoliubov Hamiltonian, which can be 
exactly diagonalized, but Hugenholtz and Pines in fact included more terms than 
these in their evaluation of the ground state energy, and hence obtain a more 
accurate result.
\end{enumerate}
It is particularly important to note that there is no basis for using $ K $
without setting $ a_0^\dagger \to \sqrt{N}$, $ a_0 \to \sqrt{N}$, as was done 
by \cite{Lewenstein You}, which can be seen as an initial attempt to get a 
Bogoliubov description which covers a range of $ N$ values.
\subsection{Formulation of the modified Bogoliubov method}
\subsubsection{Expression of the field operators in terms of phonon operators }
Our treatment will be based  on that already used for the spatially 
homogeneous case.
We therefore consider the general case for which the Hamiltonian is
\begin{eqnarray}\Label{t1}
H&=& 
-{\hbar^2\over 2m}\int d^3{\bf x}\,\psi^\dagger({\bf x})\nabla^2\psi({\bf x})
+\int d^3{\bf x}\,\psi^\dagger({\bf x})V({\bf x})\psi({\bf x})
\nonumber \\ &&
+{u\over 2}\int d^3{\bf x}\,
\psi^\dagger({\bf x})\psi^\dagger({\bf x})\psi({\bf x})\psi({\bf x}).
\end{eqnarray}
We will make the substitution
\begin{eqnarray}\Label{t2z}
\psi({\bf x}) &=&
 \left( 
a_0\xi ({\bf x}) + \sum_k \xi _k({\bf x})a_k
\right)
\\ \Label{t2}
 &\approx&
 A\left( 
\xi ({\bf x}) + {1\over \sqrt{ N}}\sum_k \xi _k({\bf x})\alpha_k
\right)
\\ \Label{t2a}
&\equiv &
 A\left( \xi({\bf x}) + {1\over \sqrt{ N}}\chi({\bf x})\right)
.
\end{eqnarray}
Here the operators $  \{a_0,a_k\}$ are independent creation and destruction 
operators satisfying the usual creation and destruction operator commutation
relations, and the set of functions $ \{\xi_0({\bf x}),\xi_k({\bf x})\}$ is a 
complete orthonormal set.  This is necessary to produce
the mandatory field operator commutation relation 
$[\psi({\bf x}),\psi^\dagger({\bf x}')] =\delta({\bf x}-{\bf x}') $.

The meaning of $ A$ and $ \alpha_k $ is essentially as previously defined 
except that the exact nature of the states involved is not yet defined.  The 
basic issue is that the occupation of the mode described by $ a_0$ is very 
large compared to the occupation of any of the modes described by the
$ a_k$.  Under these conditions, apart from the notation change
$ {\bf k}$ (momentum) $ \to k$ (an arbitrary label)  the relationship between 
the operators $ A, a_0, \alpha_k, a_k$ is exactly the same as in Sect. 
\ref{sec2}.

Notice  that $ \chi({\bf x})$ has a {\em non-local} commutation relation
\begin{eqnarray}\Label{t300}
[\chi({\bf x}),\chi^\dagger({\bf x}')]
&=& \sum_k\xi_k({\bf x})\xi^*_k({\bf x}')
\\
&=& \delta({\bf x}-{\bf x}') - \xi({\bf x})\xi^*({\bf x}')
\\
 &\equiv& R({\bf x},{\bf x}'),
\end{eqnarray}
which arises because the $ \chi$ operators act only in the subspace orthogonal 
to the condensate wavefunction $ \xi$.  These are of the same form as those for 
similar operators considered by Fetter \cite{Fetter1972}.

\subsubsection{Expansion of the Hamiltonian}
The Bogoliubov method will be valid for large $ N$ and small $ u$, and, as 
noted above, this
is expressed more precisely by setting $ u \equiv \tilde u / N $, and expanding 
the Hamiltonian in decreasing powers of $ N$ for fixed $ \tilde u$, after 
substituting for the field operators using (\ref{t2}).  
In order to eliminate $ a_0^\dagger a_0$ as in the homogeneous case, we note 
that
\begin{eqnarray}\Label{t301}
\sum_{k}\alpha_k^\dagger\alpha_k
&=& \int d^3{\bf x} \,\chi^\dagger({\bf x})\chi({\bf x}).
\end{eqnarray}
Carrying out this procedure, we then get
\begin{eqnarray}\Label{t4}
H= { N}\,{\cal H}_1+ \sqrt{ N}\,{\cal H}_2+ {\cal H}_3+ \dots
\end{eqnarray}
in which
\begin{eqnarray}\Label{t5a}
{\cal H}_1&=& 
-{\hbar^2\over 2m}\int d^3{\bf x}\,\xi^*({\bf x})\nabla^2\xi({\bf x})
+\int d^3{\bf x}\,\xi^*({\bf x})V({\bf x})\xi({\bf x})
\nonumber \\ &&
+{\tilde u\over 2}\int d^3{\bf x}\,
\big |\xi({\bf x})\big |^4,
\\ \Label{t5b}
{\cal H}_2&=& 
-{\hbar^2\over 2m}\int d^3{\bf x}
\left\{
\chi^\dagger({\bf x})\nabla^2\xi({\bf x}) +\xi^*({\bf x})\nabla^2\chi({\bf x})
\right\}
\nonumber \\
&&+\int d^3{{\bf x}}\,V({\bf x})\left\{
\chi^\dagger({\bf x})\xi({\bf x})+\xi^*({\bf x})\chi({\bf x})\right\}
\nonumber \\
&&+ {\tilde u}\int d^3{\bf x}\left\{
 |\xi({\bf x})|^2\xi({\bf x})\chi^\dagger({\bf x})
 +|\xi({\bf x})|^2\xi^*({\bf x})\chi({\bf x})\right\},
\nonumber \\
\\
\Label{t5c}
{\cal H}_3 &=&
-{\hbar^2\over 2m}\int d^3{\bf x}\,\chi^\dagger({\bf x})\nabla^2\chi({\bf x})
+\int d^3{\bf x}\,\chi^\dagger({\bf x})V({\bf x})\chi({\bf x})
\nonumber \\
&+&  \int d^3{\bf x}\bigg\{
{\tilde u\over 2}\big(\xi({\bf x})\chi^\dagger({\bf x})\big)^2 +
{\tilde u\over 2}\big(\xi^*({\bf x})\chi({\bf x})\big)^2
\nonumber \\
& +&
\chi^\dagger({\bf x})\chi({\bf x})\Big(
{2\tilde u}\big |\xi({\bf x})\big |^2- \mu\Big)\bigg\}
-{\tilde u\over 2}\int d^3{\bf y}\big|\xi({\bf y})\big|^4 .
\nonumber \\
\end{eqnarray}
in which 
\begin{eqnarray}\Label{t501}
 \mu& =& \int d^3{\bf y}
\Big(\xi^*({\bf y})\Big( -{\hbar^2\nabla^2\over 2m}\Big)\xi({\bf y}) 
\nonumber \\ && \qquad
+V({\bf y})\big|\xi({\bf y})\big |^2 +
{\tilde u} \big |\xi({\bf y})\big |^4 \Big)
\end{eqnarray}
If we now choose $ \xi$ to minimize $ {\cal H}_1$, subject to the condition 
that $ \int d^3{\bf x}\,|\xi({\bf x})|^2 =1$, we find that 
\begin{enumerate}
\item
The condition for a {\em local} minimum of $ {\cal H}_1$ is that the condensate 
wavefunction $ \xi({\bf x})$ satisfies the {\em time independent 
Gross-Pitaevskii equation}
\begin{eqnarray}\Label{t6}
-{\hbar^2\over 2m}\nabla^2\xi({\bf x}) 
+V({\bf x})\xi({\bf x}) + \tilde u \big|\xi({\bf x})\big|^2 \xi({\bf x}) 
&=& \mu \xi({\bf x}).
\nonumber\\
\end{eqnarray}
Here $ \mu$ arises as a Lagrange multiplier necessary to maintain the 
normalization of $ \xi({\bf x})$---any valid value of $ \mu$ is permitted in 
this procedure, and the value obtained is consistent with (\ref{t501})
\item  Under this condition the terms linear in $ \chi$ or $ \chi^\dagger$ 
vanish, since the 
$ \xi_{k}({\bf x})$ are a set of functions orthogonal to 
$ \xi({\bf x})$.
\end{enumerate}
Since $ N$ is known, (\ref{t6}) is to be considered as 
a nonlinear eigenvalue equation for $ \xi({\bf x})$.  Thus, the possible values 
of $ \mu$ are the eigenvalues of the Gross-Pitaevskii equation.

For stability $ {\cal H}_3$ should be positive definite;
whether this can be satisfied seems to depend on 
the actual solution of the Gross-Pitaevskii equation which is under 
consideration.  If the condition is not satisfied, the state represented by 
the solution of the Gross-Pitaevskii equation is not stable, and the extremum 
found is not a minimum.

When restricted to a fixed $ N$ subspace, the expansion 
(\ref{t4},\ref{t5a}--\ref{t5c}), is exactly the same as that obtained by Fetter 
\cite{Fetter1972} using the method based on the minimization of 
$ \langle K\rangle$.

\subsubsection{Diagonalization of ${\cal H}_3$}
The expression (\ref{t5c}) for $ {\cal H}_3 $ can be diagonalized most simply 
by working in the $ \xi_k$ basis, in much the same way as Javanainen 
\cite{Javanainen 1/97}.  This method is also preferred here since it 
generalizes rather straightforwardly to the time-dependent situation.  Thus one 
can write
\begin{eqnarray}\Label{diag1}
 {\cal H}_3 &= & E_3+
\sum_{k,q}\Bigg\{\left( L_{kq}+F_{kq}\right)\alpha^\dagger_k \alpha_q
\nonumber \\ &&
\qquad +{1\over 2} G_{kq}\alpha^\dagger_k \alpha^\dagger_q
+{1\over 2} G_{kq}^*\alpha_k \alpha_q\Bigg\}
\end{eqnarray}
where
\begin{eqnarray}\Label{diag2a}
L_{kq} &=& 
\int d^3{\bf x}\,\xi^*_k({\bf x})\left\{-{\hbar^2\over 2 m}\nabla^{2}
+V({\bf x})\right\}
\xi_q({\bf x})
\\ \Label{diag2b}
F_{kq} &=& 
\tilde u\int d^3{\bf x}\,\xi^*_k({\bf x})
\left\{2|\xi({\bf x})|^2 -\mu\right\}\xi_q({\bf x})
\\ \Label{diag2c}
G_{kq} &=& 
\tilde u\int d^3{\bf x}\,\xi^*_k({\bf x})\left\{
\xi({\bf x})^2\right\}\xi_q({\bf x})
\\ \Label{diag2d}
E_3 &=& -{\tilde u\over 2}\int d^3{\bf y}\,|\xi({\bf y})|^4
\end{eqnarray}
\subsubsection{Quasiparticles}
The Hamiltonian $ {\cal H}_3$ can be diagonalized by a Bogoliubov
transformation of the form
\begin{eqnarray}\Label{tb15}
\alpha_k = \sum_mc_{km}b_m + \sum_ms_{km}b^\dagger_m
\end{eqnarray}
and here $ b_m$ is a quasiparticle destruction operator.  We can then 
write
\begin{eqnarray}\Label{tb16}
\chi({\bf x}) = 
\sum_m\left(p_m({\bf x})b_m + q_m({\bf x})b^\dagger_m\right)
\end{eqnarray}
with
\begin{eqnarray}\Label{tb17}
p_m({\bf x}) &=& \sum_k c_{km}\xi_k({\bf x})
\\ \Label{ tb18}
q_m({\bf x}) &=& \sum_k s_{km}\xi_k({\bf x}).
\end{eqnarray}
The diagonalized Hamiltonian is then written
\begin{eqnarray}\Label{tb19}
{\cal H}_3
 = \hbar\omega_g(N) + \sum_m\hbar\epsilon_m(N) b^\dagger_mb_m.
\end{eqnarray}

Notice that, even though the $ N$ dependence is not always explicitly written, 
almost everything in the above is a function of $ N$.

\subsection{Relationship between ground states for $ N$ and $ N+1$ 
particles}
The operator $ A$ contains an $ N$ dependence which arises from the change in 
the shape of the ground state wavefunction $ \xi({\bf x})$ as $ N$ changes.  
This is an effect which does not arise in a spatially homogeneous situation.
Thus we can exhibit this feature by writing, as a result of (\ref{t2z}),
\begin{eqnarray}\Label{cor1}
a_0(N) = \int d^3{\bf x}\,\xi^*({\bf x},N)\psi({\bf x}),
\end{eqnarray}
where the explicit dependence on $ N$ of both $ \xi$ and $ a_0$ has now been 
written. 

This means that $ a_0^\dagger(N)|N, {\bf n} = {\bf 0}\rangle$
is a state with $ N+1$ particles in the wavefunction corresponding to ground 
state of the
$ N$ particle state; it is not the ground state for $ N+1$ particles.  In other 
words, $ a_0^\dagger(N)$ is not the operator which converts an $ N $ particle 
ground state into an $ N+1$ particle ground state.

We will therefore compute the appropriate operator.
We can write
\begin{eqnarray}\Label{cor2}
a_0(N+1) &=& a_0(N)
\nonumber \\
& +& \int d^3{\bf x}\Big(
\xi^*({\bf x},N+1) - \xi^*({\bf x},N)\Big)\psi({\bf x})
\\ \Label{cor3}
&\approx &  a_0(N) + 
\int d^3{\bf x}{{\partial \xi^*({\bf x},N) \over\partial N }}\psi({\bf x}).
\end{eqnarray}
Now expand $ {\partial \xi^*({\bf x},N)/\partial N }$ in the $ N$ particle 
basis states as 
\begin{eqnarray}\Label{cor4}
{\partial \xi^*({\bf x},N) \over\partial N } &=&
{1\over N}\left(i r_0\xi^*({\bf x},N) 
+\sum_k r_k \xi_k^*({\bf x},N)\right)
\end{eqnarray}
The requirement that $ \xi$ be normalized allows us to deduce that 
$ r_0 $ is real, and by redefining
 $ \xi(N) \to  \xi(N)\exp\{-i\int dN\, r_0(N)/N \}$
we can transform $ r_0$ to zero; that is, we can choose a family of solutions 
of the time-independent Gross-Pitaevskii equation in which the relative phases
of the members of the family are such that $ r_0=0$.  Once $ r_0$ has been 
eliminated, it is clear that $ r_k$ are of order of magnitude 
$ 1$.  This means that we can write approximately
\begin{eqnarray}\Label{cor5}
a_0(N+1) \approx a_0(N) + {1\over N}\sum_k r_k a_k(N).
\end{eqnarray}
We now want to relate the $ N$ particle ground state to the $ N+1 $ particle 
ground state.  We can write
\begin{eqnarray}\Label{cor501}
|N\rangle_N  &
\equiv&{\left\{a_0^\dagger(N)\right\}^N\over\sqrt{N!}}|0\rangle
\\ \Label{cor502}
|N+1\rangle_{N+1}  &
\equiv&{\left\{a_0^\dagger({N+1})\right\}^{N+1}\over\sqrt{{(N+1)}!}}|0\rangle .
\end{eqnarray}
We first write (\ref{cor5}) in terms of the phonon operators $ \alpha_k(N)$
thus
\begin{eqnarray}\Label{cor6}
a_0(N+1) = 
a_0(N)\left(1 + \sum_k {r_k\over N}{ \alpha_k(N)\over\sqrt{N}}\right).
\end{eqnarray}
and raise this to the power $ N+1$, using the binomial theorem to get, 
accurate 
to order $ 1/N$,
\begin{eqnarray}\Label{cor7}
\left\{a_0^\dagger({N+1})\right\}^{N+1} &\approx&
 \left(1+\sum_k {r^*_k \alpha^\dagger_k(N)\over\sqrt{N}}\right)
\left\{a_0^\dagger(N)\right\}^{N+1}.
\nonumber \\
\end{eqnarray}
From (\ref{cor501},\ref{cor502}) this means that 
\begin{eqnarray}\Label{cor8}
|N+1\rangle_{N+1} 
&=&{\Big\{a_0^\dagger({N+1})\Big\}^{N+1}\over\sqrt{{(N+1)}!}} 
   {\Big\{a_0(N)\Big\}^N\over\sqrt{N!}}
    |0\rangle
\\ \Label{cor9}
&\approx&
\left(1+ {\sum_kr^*_k\alpha^\dagger_k\over\sqrt{N}}\right)
{a^\dagger_0(N)\over\sqrt{N+1}}|N\rangle_N
\end{eqnarray}
Similar procedures can be carried out for the $ a_k(N)$ and hence the
 $ \alpha_k(N)$, but the differences between these at $ N$ and $ N+1$ will have 
a negligible effect because of their small occupation numbers.

From the above, we can say that the operator $ B^\dagger(N)$ which connects the 
$ N$ and $ N+1$ ground states through
\begin{eqnarray}\Label{cor901}
B^\dagger(N)|N\rangle_N =  \sqrt{N+1} |N+1\rangle_{N+1}
\end{eqnarray}
is given  approximately by (to order $ 1/N$)
\begin{eqnarray}\Label{cor10}
B^\dagger(N) &\approx &
\left(1+{1\over\sqrt{N}}\sum_kr^*_k\alpha^\dagger_k\right)a_0^\dagger(N).
\end{eqnarray}
This means that the field operator expansion (\ref{t2a}) now takes the form
\begin{eqnarray}\Label{cor11}
\psi({\bf x}) &\approx&
B(N)\left(\xi({\bf x}) 
+{1\over\sqrt{N}}\chi_B({\bf x})\right),
\end{eqnarray}
where
\begin{eqnarray}\Label{cor1101}
\chi_B({\bf x}) &\equiv &\chi({\bf x}) - \xi({\bf x})\sum_k r_k\alpha_k
\\ &=& \sum_m\left( f_m({\bf x}) b_m + g_m({\bf x}) b^\dagger_m\right)
\end{eqnarray}
and
\begin{eqnarray}\Label{cor1102}
f_m({\bf x}) &=& p_m({\bf x}) - \xi({\bf x})\sum_{m}r_kc_{km}
\\
g_m({\bf x}) &=& q_m({\bf x}) - \xi({\bf x})\sum_{m}r_ks_{km}
\end{eqnarray}
The physical distinction between the quasiparticle wavefunctions $ p_m, q_m$ 
and the modified wavefunctions $ f_m, g_m$ is that the first set are amplitudes 
for the production or absorption  of quasiparticle without changing $ N$, 
{\em e.g.}, by application of sound waves, whereas the second set give the 
corresponding amplitudes for changing quasiparticle numbers by changing the 
particle number by 1.

It is particularly interesting to see that the corrections to the 
quasiparticle term are of the same order of magnitude as the original terms;
the correction is thus very significant.  It reflects the fact that the change 
in the groundstate wavefunction from $ N$ to $ N+1$ particles affects the $ N$ 
particles already present as well as the added particle.

\section{Time-dependent solutions}\Label{sec4} 
Suppose we now consider a representation like (\ref{t2z}--\ref{t2a}) in which 
however we have a time-dependent condensate  wavefunction $ \xi({\bf x},t) $.
This is with fixed $ N$ since we are not considering condensate growth; rather, 
we are investigating the situation where the condensate has been 
macroscopically disturbed from the stationary state as in the recent 
experiments in JILA\cite{JILA osc} and MIT\cite{MIT osc}.

Let us proceed within the Schr\"odinger picture as follows.  Firstly suppose 
the condensate wavefunction (normalized to 1) is $ \xi({\bf x},t)$ and define 
condensate and non-condensate parts of the field operator by
\begin{eqnarray}\Label{td1}
\psi_{0}({\bf x},t) &=& 
\xi({\bf x},t)\int d^3{\bf x'}\,\xi^*({\bf x}',t)\psi({\bf x}') 
\\
&\equiv & a_{0}(t)\xi({\bf x},t)
\\
\psi_{\rm nc}({\bf x},t) &=& \psi({\bf x}) - \psi_{0}({\bf x},t) .
\end{eqnarray}
We can set up non-condensate mode operators by defining
\begin{eqnarray}\Label{td4}
a_k(t) &=& \int d^3{\bf x}\,\xi^*_k({\bf x},t)\psi({\bf x}) 
\end{eqnarray}
where $ \{\xi_k({\bf x},t)\} $ are any orthonormal set of wavefunctions 
orthogonal to the condensate wavefunction $ \xi({\bf x},t)$.

The total number operator is
\begin{eqnarray}\Label{td2}
{\cal N} &=& \int d^3{\bf x}\,\psi^\dagger({\bf x}) \psi({\bf x}) 
\end{eqnarray}
and is time independent.
We can correspondingly define condensate and non-condensate number operators 
by
\begin{eqnarray}\Label{td3}
{\cal N}_{0}(t) &=& a^\dagger_{\rm nc}(t) a_{\rm nc}(t) \\
{\cal N}_{\rm nc}(t) &=& {\cal N} - {\cal N}_{0}(t)
\\ &=& 
\int d^3{\bf x}\,\psi_{\rm nc}^\dagger({\bf x},t) \psi_{\rm nc}({\bf x},t).
\end{eqnarray}
The time-dependence of these operators arises because their definitions, which 
involve the time-dependent wavefunctions $ \xi({\bf x},t),\xi_k({\bf x},t)$, 
change with time; they are however still operators in the Schr\"odinger 
picture.

Using these definitions it is possible to define the operator $ A $ essentially 
as before so that,
\begin{eqnarray}\Label{td5}
a_{0}(t)  &=&
 \sqrt{1-{{{ N}_{\rm nc}(t)\over { N}}}}\,\, A(t).  
\end{eqnarray}
Although $ A(t)$ is time-dependent, $  {\cal N} = A^\dagger(t)A(t)$ is time 
independent.  The time-dependence of $ A(t)$ comes about because of the
changing definition of the modes unaffected by its action.  That is, $ A$ 
reduces the total number of particles while leaving the number of 
non-condensate particles the same, but the definitions of the condensate and 
non-condensate modes are themselves time dependent.
Further, although the 
operators for condensate and non-condensate modes commute, the $ A(t)$ operator 
does not commute exactly with any of $ a_{0}(t)$, $ a^\dagger_{0}(t)$, 
$ a_{k}(t)$ or $ a^\dagger_{k}(t)$.

It will be particularly important to keep track of all time-dependences which 
are a result of the the expression of the time-independent Schr\"odinger 
picture field operators $ \psi({\bf x})$ as projections on the time-dependent 
basis vectors $ \{\xi({\bf x},t),\xi_k({\bf x},t)\}$---the only operator in 
this basis that does not develop such a time-dependence is the operator 
$ {\cal N}$.  We will show that a unitary transformation can be introduced 
which 
transforms us to a kind of interaction picture, which we shall call the 
{\em condensate picture}, in which we have a description in terms of 
{\em time-independent} phonon creation and destruction operators 
$ \beta^\dagger_k, \beta_k$ corresponding to the time-dependent modes 
$ \xi_k({\bf x},t) $ provided that {\em the condensate wavefunction 
$ \xi({\bf x},t)$ satisfies the time-dependent Gross-Pitaevskii equation}
\begin{eqnarray}\Label{td11}
i\hbar{\partial\xi \over\partial t } 
= -{\hbar^2\over 2m}\nabla^2\xi +V\xi+ \tilde u|\xi |^2\xi + \bar\mu(t) \xi
\end{eqnarray}
where $ \bar\mu(t) $ is arbitrary, and of course can be eliminated by 
multiplying $ \xi$ by an appropriate time-dependent phase.
\subsection{Time dependence of the phonon operators}
\subsubsection{Explicit time-dependence}
Since $ \{\xi({\bf x},t),\xi_k({\bf x},t)\}$ are a complete orthonormal set, we 
can define a projector onto the non-condensate modes by
\begin{eqnarray}\Label{td12}
R({\bf x},{\bf x}',t) &=& \sum_k\xi_k({\bf x},t)\xi^*_k({\bf x}',t)
\\ \Label{td13}
&=& \delta({\bf x}-{\bf x}') - \xi({\bf x},t)\xi^*({\bf x}',t)
\end{eqnarray}
so that, using a time-dependent version of (\ref{t2z}--\ref{t2a}),
\begin{eqnarray}\Label{td14}
A(t){1\over\sqrt{ N}}\chi({\bf x},t) \approx
\int d^3{\bf x}'\,R({\bf x},{\bf x}',t)\psi({\bf x}')
\end{eqnarray}
and this means that to the lowest order in $\sqrt{N} $ we can write
\begin{eqnarray}\Label{td15}
\chi({\bf x},t) =
{1\over\sqrt{ N}}A^\dagger(t)\int d^3{\bf x}'\,R({\bf x},{\bf x}',t)\psi({
\bf x}')
\end{eqnarray}
The explicit time-dependence of $ \chi$ arises from the projector and the 
operator $ A^\dagger(t)$.  We use expression
\begin{eqnarray}\Label{td16}
 A^\dagger(t)\approx a^\dagger_{0}(t)
 =\int d^3{\bf x}\,\xi({\bf x},t)\psi^\dagger({\bf x})
\end{eqnarray}
so that
\begin{eqnarray}\Label{td17}
{\partial A^\dagger(t) \over\partial t } &\approx& 
\int d^3{\bf x}\,\dot\xi({\bf x},t)\psi^\dagger({\bf x})
\\
&\approx & A^\dagger(t)\int d^3{\bf x}\,\dot\xi({\bf x},t)\xi^*({\bf x},t).
\end{eqnarray}
Now using the explicit form (\ref{td13}) for the projector $ R$, and 
resubstituting for $ \psi$ in terms of $ A$ and $ \chi$ we deduce
\begin{eqnarray}
{{\partial\chi({\bf x},t) \over\partial t}}& =&
\left\{
\int d^3{\bf x}'\dot\xi({\bf x}',t)\xi^*({\bf x}',t)
\right\}\chi({\bf x},t)
\nonumber \\ 
\Label{td18}
&+ &\int d^3{\bf x}' \dot R({\bf x},{\bf x}',t)
\left\{\sqrt{ N}\,\xi({\bf x}',t) +\chi({\bf x}',t)\right\}
\nonumber \\
\end{eqnarray}
This time-dependence is only the {\em explicit} time-dependence, and does not 
include that arising from the commutator with the Hamiltonian (\ref{t4}). 

\subsubsection{Expansion of the Hamiltonian and the condensate picture}
The Hamiltonian can now be expansied in decreasing powers of $ \sqrt{N}$ in 
exactly the same way as for the time independent case, with appropriate 
substitutions.  However, the   largest terms arising from  (\ref{t5b}) and  
(\ref{td18})  are both proportional to $ \sqrt{N}$, and these must 
be arranged to cancel, since otherwise $ \chi/\sqrt{ N}$ would develop a 
c-number component  comparable with $ \xi$, invalidating the expansion (
\ref{t2a}).  We will show that this cancellation is only possible if 
$ \xi({\bf x},t)$ satisfies the time-dependent Gross-Pitaevskii equation.

The coefficient of $ \sqrt{ N}$ in (\ref{td18})  can be written as
\begin{eqnarray}\Label{td19}
 \int d^3{\bf x}' \dot R({\bf x},{\bf x}',t)
\xi({\bf x}',t) = 
{i\over\hbar}[{\cal H}_R(t),\chi({\bf x},t)]
\end{eqnarray}
in which 
\begin{eqnarray}\Label{td20}
{\cal H}_R(t)& = &i\hbar\int d^3{\bf x}\left\{
\dot\xi({\bf x},t)\chi^\dagger({\bf x},t)- \dot\xi^*({\bf x},t)\chi({\bf x},t)
\right\}.
\nonumber\\
\end{eqnarray}
Here we have used the identity 
\begin{eqnarray}\Label{td201}
0&=& \int d^3{\bf x}'\dot R({\bf x},{\bf x}',t)\xi({\bf x}',t)
+\int d^3{\bf x}' R({\bf x},{\bf x}',t)\dot\xi({\bf x}',t),
\nonumber \\
\end{eqnarray}
which arises from 
the fact that $ \int d^3{\bf x}' R({\bf x},{\bf x}',t)\xi({\bf x}',t)=0$.)  

If $ \xi$ satisfies the time-dependent Gross-Pitaevskii equation 
(\ref{td11}), it is clear that $ {\cal H}_R =-{\cal H}_2  $. This means that
we can make a unitary transformation from the Schr\"odinger picture to a 
picture which we call the {\em condensate picture}, defined by
\begin{eqnarray}\Label{td21}
|\Phi,t\rangle &\to & V(t)|\Phi,t\rangle\equiv |\Phi,t\rangle_c
\\ \Label{td22}
\psi({\bf x}) &\to & V(t)\psi({\bf x})V^{-1}(t) \equiv\psi_c({\bf x},t)
\\ \Label{td22a}
{d \over dt}V(t)& =& -{i\over\hbar}\sqrt{ N}\,{\cal H}_R(t) V(t).
\end{eqnarray}
In this picture
the part of the explicit time-dependence of $ \chi$ proportional to
 $ \sqrt{ N}$  in (\ref{td18}) will disappear, and the term 
$ \sqrt{ N}\,{  H}_2$ in the equation of motion for the states 
$|\Phi,t\rangle_c $ will cancel with a term arising from the unitary 
transformation.  We will then be able to write the equation of motion  in this 
picture as (to order $ N^0$)
\begin{eqnarray}\Label{td23}
\left({ N}\,{\cal H}_1+  {\cal H}_3(t)\right)|\Phi,t\rangle_c
=i\hbar {d\over dt}|\Phi,t\rangle_c
\end{eqnarray}
in which
\begin{eqnarray}\Label{td24}
&&{\cal H}_1 =
-{\hbar^2\over 2m}\int d^3{\bf x}\,\xi^*({\bf x},t)\nabla^2\xi({\bf x},t)
\nonumber \\ 
&&\quad
+\int d^3{\bf x}\,\xi^*({\bf x},t)V({\bf x})\xi({\bf x},t)
+{\tilde u\over 2}\int d^3{\bf x}\,
\big |\xi({\bf x},t)\big |^4,
\\
\Label{td25}
&&{\cal H}_3(t) =\int d^3{\bf x}\Bigg[-{\hbar^2\over 2m}
\chi_{c}^\dagger({\bf x},t)\nabla^2\chi_{c}({\bf x},t)
\nonumber \\
&&\quad
+\chi_{c}^\dagger({\bf x},t)V({\bf x})\chi_{c}({\bf x},t)
\nonumber \\
&&\quad
+{\tilde u\over 2} \big(\xi({\bf x},t)\chi_{c}^\dagger({\bf x},t)\big)^2 +
{\tilde u\over 2}\big(\xi^*({\bf x},t)\chi_{c}({\bf x},t)\big)^2
\nonumber \\
&&\quad
+\chi_{c}^\dagger({\bf x},t)\chi_{c}({\bf x},t)\big[
2\tilde u \big |\xi({\bf x},t)\big |^2
 -\mu(t)\big]
\Bigg]
\nonumber \\ && \qquad
-{\tilde u\over 2}\int d^3{\bf y}\big|\xi({\bf y},t)\big|^4 .
\end{eqnarray}
in which $ \mu(t)$ has no connection with the $ \bar \mu(t)$ in (\ref{td11}), 
but is given by
\begin{eqnarray}\Label{td25001}
 \mu(t)& =& \int d^3{\bf y}
\Big(\xi^*({\bf y},t)\Big( -{\hbar^2\nabla^2\over 2m}\Big)\xi({\bf y},t) 
\nonumber \\ &&\qquad
+V({\bf y})\big|\xi({\bf y},t)\big |^2  +
{\tilde u} \big |\xi({\bf y},t)\big |^4 \Big)
\\
&=& -i\hbar
\int d^3{\bf y}\xi^*({\bf y},t){\partial \xi({\bf y},t) \over\partial t }
\end{eqnarray}
Notice that no time-dependence is written for $ {\cal H}_1$, since 
it is in fact a constant c-number when $ \xi$ satisfies the time-dependent 
Gross-Pitaevskii equation.

Note that although the equation of motion is given by (\ref{td23}), the 
{\em energy} is given by the full Hamiltonian 
$ H = { N}\,{\cal H}_1+  \sqrt{ N}\,{\cal H}_2+  {\cal H}_3(t)$, with 
the 
operators evaluated in the condensate picture.  Since there is no 
time-dependence of the Schr\"odinger picture Hamiltonian, the total energy must 
be conserved.

\subsubsection{Time-independent phonon operators}
The expansion of the phonon field in the condensate picture as
\begin{eqnarray}\Label{td26}
\chi_{c}({\bf x},t) = \sum_k\xi_k({\bf x},t)\alpha_k
\end{eqnarray}
does not automatically require that $ \alpha_k$ be time independent.  However 
the mode functions $ \xi_k$ have been so far essentially arbitrary.  The 
explicit time-dependence of $ \chi_{c}$ in the condensate picture is given by 
omitting the term proportional to $ \sqrt{N}$ in (\ref{td18}), and by 
inserting 
the expansion (\ref{td26}) into this, we find that $ \alpha_k$ can be chosen to 
be independent of time if the mode functions satisfy the equation of motion
\begin{eqnarray}\Label{td27}
{\partial\xi_k({\bf x},t) \over\partial t} &=&
\xi_k({\bf x},t) 
\left\{\int d^3 {\bf y}\,\xi^*({\bf y},t) \dot\xi({\bf y},t) \right\}
\nonumber \\ 
&& -\xi({\bf x},t) 
\left\{\int d^3 {\bf y}\,\dot\xi^*({\bf y},t) \xi_k({\bf y},t) \right\}
\end{eqnarray}
It is straightforward to check that the requirements that $ \xi_k$ form an 
orthonormal set, and are orthogonal to $ \xi$ are both preserved by this 
equation of motion.  

Substituting the expansion (\ref{td26}) into $ {\cal H}_3(t)$ as given by 
(\ref{td25}) we obtain $ {\cal H}_3(t)$ in the form
\begin{eqnarray}\Label{td28}
 {\cal H}_3(t) &= & E_3(t) + \sum_{k,q}\bigg\{
\left(L_{kq}+ F_{kq}(t)\right)\alpha^\dagger_k \alpha_q
\nonumber \\
&&+ G_{kq}(t)\alpha^\dagger_k \alpha^\dagger_q
+ G_{kq}^*(t)\alpha_k \alpha_q\bigg\}.
\end{eqnarray}
Here $  L_{kq}(t)$, $  F_{kq}(t)$, $  G_{kq}(t) $ and $E_3(t) $ are defined in 
the same way as in (\ref{diag2a}--\ref{diag2d}), but using in this case the 
condensate picture operators and the time-dependent mode functions which 
are now arbitrary only at the initial time, since the 
equation of motion (\ref{td27}) gives them for all future times.

\subsection{Summary of the time-dependent description}
We have shown that the field operators can be written in a time-dependent 
version of the form (\ref{t2a}).  This expansion is  valid as an asymptotic 
expansion in $ 1/\sqrt N$ if and only if $ \xi({\bf x},t)$ satisfied the time 
dependent Gross-Pitaevskii equation.

\section{Conclusion}
The adapted Bogoliubov method presented in this paper gives a precise meaning 
to the definition of the macroscopic wavefunction usually 
defined by the limiting procedure \cite{LandP} 
\begin{eqnarray}\Label{td29}
\langle N,m,t|\psi^\dagger({\bf x})|N-1,m,t\rangle 
\sim\sqrt{N}\,\xi^*({\bf x},t),
\end{eqnarray}
where, in the words of \cite{LandP}, 
\begin{quote}``...since the condensate contains a 
macroscopically large number of particles, changing this number by 1 does not 
essentially affect the state of the system; we may say that the result of 
adding 
(or removing) one particle in the condensate is to convert a state of the 
system of $ N $ particles into the `same' state of a system of $ N+1 $ 
particles.  ... the symbols $ |N,m,t\rangle $ and $ |N+1,m,t\rangle $ denote 
two `like' states differing only as regards the number of particles in the 
system.''
\end{quote}
This definition has the disadvantage of being rather vague about 
exactly what is meant by the concept of the `same' or `like' states 
which differ only in that the values of $ N$ differ by 1.  From this 
paper it is clear that the description of the eigenfunctions of the 
non-condensed particles depends on the value of $ N $, and although 
the difference between the eigenfunctions for $ N$ and $ N+1$ must 
become negligible for $ N\to \infty$, if we are contemplating a 
situation in which the condensate is growing, as in recent 
experiments, this difference may have non-negligible effects.

Our method can also be seen as the logical completion of the aim of Hugenholtz 
and Pines \cite{Hugen} to treat the Bose condensed gas by elimination the 
ground state.  Their method was only able to do this at the expense of breaking 
the exact conservation of particle numbers, whereas our method puts the 
approximation in the relationship between  the operators $ A, \alpha_k$ and the 
particle operators, and in their exact commutation relations.  However, there 
is no good reason to why one should not in principle seek a more accurate 
approximation than has been done here by computing terms of higher 
order in the inverse $\sqrt{N}$ expansion, and get results comparable with 
those 
of Hugenholtz and Pines.  For the condensates at present in existence this is 
not 
an urgent problem; rather the main problem is to get as simple a description as 
possible of the eigenfunctions as a function of $ N$ as well as the other 
variables.  

Griffin\cite{Griffin} has recently shown how the 
Hartree-Fock-Bogoliubov method can also be used.  This is closely 
related to any Bogoliubov method, including this one, but it suffers 
from not being a systematic expansion in any parameter.  For example, 
depending on assumptions made on certain averages one may or may not 
obtain a gapless spectrum.  A systematic method would identify all 
terms of a given order in an appropriate small parameter, and either 
use all of them or use none of them.  If the Hartree-Fock-Bogoliubov 
method is developed in inverse powers of $\sqrt{N}$, one will find the 
same results as presented here to the degree of accuracy presented 
here.

It is well known that the Bogoliubov method is {\em 
gapless}\cite{Hohenberg Martin,Griffin} (in the long wavelength limit, 
the energy levels approach the energy of the ground state; i.e., a 
phonon of very long wavelength has vanishingly small energy).  The 
method of derivation used here shows that the non-conservation of 
particle numbers, often seen as the hallmark of the method, is not at 
all essential to the method; that it is really takes only a slightly 
different point of view to see that the non conservation of particle 
numbers arises from a slightly inappropriate way of looking at the 
quasiparticle operators.  Nevertheless, the elimination of this 
problem is absolutely essential if we wish to study condensate growth.  
Approximations which do not preserve particle number conservation can 
give rise to spurious terms in the equations of motion for the 
development of a condensate, and thus make it difficult to identify 
the true details of the growth process.

The reason for developing the modified Bogoliubov method is thus to 
apply the method to the growth of the condensate, and this will 
require its incorporation into the framework of Quantum Kinetic Theory 
\cite{QKI}.  If the condensate growth is rather slow, it will adequate 
to use the time-independent formalism of Sect.\ref{sec3}, but for 
faster condensate growth, the time-dependent formalism will be 
indispensable.  These aspects will be treated in \cite{QKII}.
\acknowledgements
I would like to thank Rob Ballagh and Keith Burnett for helpful comments.
This work was supported by the Marsden Fund under contract number
PVT-603.  

%\end{multicols}
\end{document}